\documentclass[preprint]{aastex}

\tighten
\newcommand{\asec}{\hbox to 1pt{}\rlap{$^{\prime\prime}$}.\hbox to 2pt{}}
\newcommand{\amin}{\hbox to 1pt{}\rlap{$^{\prime}$}.\hbox to 2pt{}}

\shortauthors{Lauer et al.}
\shorttitle{M31 Nuclear Blue Stars}

\begin{document}

\title{The Cluster of Blue Stars Surrounding the M31 Nuclear Black Hole
\footnote{Based on observations made with the NASA/ESA
{\it Hubble Space Telescope}, obtained at the Space Telescope Science Institute,
which is operated by the Association of Universities for
Research in Astronomy, Inc., under NASA contract NAS 5-26555. These
observations are associated with GO \# 10572.}}

\author{Tod R. Lauer,\altaffilmark{1} Ralf Bender,\altaffilmark{2, 3}
John Kormendy,\altaffilmark{4,2,3} Philip Rosenfield,\altaffilmark{5}
\and Richard F. Green\altaffilmark{6}}
\altaffiltext{1}{National Optical Astronomy Observatory\footnote{The
National Optical Astronomy Observatory is operated by AURA, Inc.,
under cooperative agreement with the National Science Foundation.},       
P.O. Box 26732, Tucson, AZ 85726}
\altaffiltext{2}{Universit\"ats-Sternwarte  M\"unchen, Scheinerstrasse 1,
M\"unchen D-81679, Germany}
\altaffiltext{3}{Max-Planck-Institut f\"ur Extraterrestrische Physik, Giessenbachstrasse, D-85748 Garching, Germany}
\altaffiltext{4}{Department of Astronomy, University of Texas at Austin, 1 University Station C1400, Austin, TX 78712-0259, USA}
\altaffiltext{5}{Astronomy Department, University of Washington, Seattle, WA 98195}
\altaffiltext{6}{Large Binocular Telescope Observatory, University of Arizona, 933 North Cherry Avenue, Tucson, AZ 85721, USA}

\vfill

\begin{abstract}

We obtained $\rm{U_{330}}$ and B band images of the M31 nucleus using the
High Resolution Camera of the Advanced Camera for Surveys on board the Hubble
Space Telescope (HST).
The spatial resolution in the $\rm{U_{330}}$-band, $0\asec03$ FWHM,
or 0.1 pc at M31, is sufficient to resolve the outskirts of the
compact cluster (P3) of UV-bright stars surrounding the M31 black hole.
The center of the cluster is marked by an extended source that is both
brighter and redder than the other point sources within P3; it is likely
to be a blend of several bright stars.
We hypothesize that it marks the location of the M31 black hole.
Both stellar photometry and a surface brightness fluctuation analysis,
show that the P3 stellar population is consistent with
early-type main sequence stars formed in
a $\sim100~-~200$ Myr old starburst population.
Evolutionary tracks of post early
asymptotic giant-branch stars, associated with
late-stage evolution of an old population, also traverse the U and
${\rm U-B}$ domain occupied by the P3 stars; but we argue that
only a few stars could be accounted for that way.
PEAGB evolution is very rapid, and there is no progenitor
population of red giants associated with P3.
The result that P3 comprises young stars is consistent with inferences from
earlier HST observations of the integrated light of the cluster.
Like the Milky Way, M31 harbors a
black hole closely surrounded by apparently young stars.

\end{abstract}


\section{Blue Stars / Black Hole}

There is a highly-compact cluster of blue stars \citep{l98, brown}
at the heart of the M31 double nuclear-cluster of old stars.
\citet{kb} showed that this blue cluster in turn hosts the $\sim10^8M_\odot$
super-massive black hole also known to reside in the
nucleus \citep{d84, dr, k88, r90}.  \citet{bend} argue that
the integrated spectrum and spectral-energy distribution of
the stars are consistent with their formation in a burst 200 Myr ago.
Young stars are also closely bound to the nuclear black hole in our galaxy
\citep{forrest, allen, krabbe},
suggesting that this may not be a rare phenomenon.
Understanding how apparently young stars were formed deeply interior to
the tidal field of a supermassive black hole
probes several issues of how stars and gas interact within
such an extreme environment (see \citealt{alex} for a review).
Study of the blue cluster in M31 provides an additional context
in which to test theories for the formation of the
unusual population surrounding the Milky Way black hole.

There is a diversity of ideas for explaining the origin
of the M31 blue cluster.
\citet{chang}, for example, argue that the cluster is
the expected consequence of the non-axisymmetric structure
of the surrounding M31 nucleus funneling gas from stellar mass-loss into orbit
around the black hole, where it periodically reaches surface
densities sufficient to induce collapse and star formation.
Likewise, \citep{levin}, \citet{br}, and \citet{ward}
have advanced the formation
of compact disks of young stars around the Milky Way black hole by
massive central accretion of gas.
In contrast, \citet{demar} argue that the blue stars in M31
result from stellar collisions and tidal stripping acting on
an old population of stars interacting within the high-velocity
orbits and strong tidal-field associated with the black hole.
We present new observations of the blue
cluster obtained at the maximum angular resolution offered
by the {\it Hubble Space Telescope} to provide better
constraints on the origin of the stars residing in this
unusual environment.

\subsection{The Discovery of the Blue Cluster}

The discovery of the blue cluster proceeded in stages.
\citet{king} imaged the nucleus
of M31 with the HST Faint Object Camera at 1750\AA, seeing a double
structure similar to that discovered in the optical by \citet{l93}.
By the simple expedient of checking the position-angle of their images, however,
they discovered that the optically-dimmer peak,
designated P2 by Lauer et al., was actually brighter in the UV than P1.
\citet{king} lacked $S/N$ sufficient to resolve
the spatial structure of the UV source at P2,
noting only that it was highly-compact
and possibly consistent with a point source, such as low-level emission
associated with the weak AGN detected in the radio \citep{crane},
or emission from a single PAGB star, such as those seen further out from
the nucleus \citep{k92,bert}.  In retrospect, the discovery
of a UV-bright nuclear source explains the observations of \citet{nieto},
who found the nucleus to have color gradient becoming {\it bluer}
towards its center, using CFHT images obtained at 3750\AA.
Discovery of the cluster may also have been presaged
by an enigmatic reference in \citet{red}
to third-party unpublished ``Mt. Wilson material'' purporting to show that
``the spectrum of the nucleus is of a peculiar dwarf
A type,'' a result supported by no other work prior to \citet{bend}
to the best of our literature research.\footnote{We have also been
unable to find photographic spectra of the M31 nucleus
prior to 1937 in catalogues of Mt.\ Wilson observations.  We thank
Dr. Christopher Burns of the Carnegie Observatories for searching for this
material.}

Later HST observations by \citet{l98} and \citet{brown} were able to
resolve the UV-source; both papers argued that it is a cluster of stars.
WFPC2/PC images at 3000\AA\ obtained by \citet{l98} showed that
the half-power radius of the blue source is $\sim0.2$ pc.
\citet{l98} further combined their U and V band fluxes of the source
with the \citet{king} fluxes at 1750\AA\ 
to conclude that the source is consistent with an A-star spectral
energy distribution.

\citet{bend} refined this picture further by obtaining HST STIS spectroscopy
of the nucleus over $\lambda\approx3600-5100$\AA,
and reanalyzing the F300W images presented in \citet{l98}.
The spectroscopy shows strong Balmer lines, including
a strong Balmer break, consistent with either A0 main-sequence or giant
stars, but not a population of white dwarfs of the same photospheric
temperature.  The best match to the spectrum is provided by a $200\pm50$ Myr
population formed in a single burst.
Populations of less than half this age would exhibit
too much UV-flux to the blue side of the Balmer break.

\citet{bend} further show that the stars are distributed
in a flat disk with an exponential profile of scale-length of $0.37\pm0.04$ pc
in rapid rotation about the black hole.\footnote{We
use 770 Kpc as the distance to M31 \citep{dm31}.}
The rotational broadening of the STIS spectrum implies a black hole mass of
$M_\bullet=1.4\times10^8M_\odot,$ a significant upwards revision from
earlier determinations.

\subsection{The Blue Cluster in the Context of the M31 Nucleus}

The M31 black hole and its cluster of blue stars mark the center of a much
larger nuclear star-cluster of complex structure.
The surface brightness of the nuclear stellar system  begins to rise above
the underlying M31 bulge interior to $r\sim3''$ or $\sim10$ pc from the center.
The cluster becomes increasingly elliptical at smaller radii, with
$1-b/a >0.3$ for $r<1\asec7,$ but its isophotes remain concentric with
the photo-center of the bulge for $r>1\asec4$ \citep{l93}.
At yet smaller radii the nuclear cluster exhibits the
double-peaked structure discovered by \citet{l93}.
The optically-brighter peak, P1, is offset by $0\asec49$ from
the secondary brightness peak, P2,
and a slightly lesser amount from the M31 bulge photo-center.

The best explanation for the double morphology is that both peaks
result from the line-of-sight projection of an eccentric disk of stars
roughly $\sim2$ pc in diameter bound to the black hole \citep{trem}.
The black hole strongly dominates the potential, thus the stars
in the disk follow Keplerian orbits,
spending most of their time at the orbital apo-center, creating
an enhancement of light at P1.
Significantly, the P1---P2 line twists away from the major axis
of the outer nucleus by $\sim20^\circ.$
P1 and P2 are both redder in $V-I$ than the surrounding
nucleus \citep{l98}.
High spatial-resolution spectroscopy shows that both P1 and P2
comprise old stellar populations with characteristics
more like each other, than the underlying bulge \citep{kb}.
The ``Tremaine Disk'' thus appears to
be a distinct component embedded in a much larger nuclear-cluster.
The initial model of the disk provided by \citet{trem} has been
refined by the more detailed analysis presented by \citet{pt03}.
An interesting feature of the refined model is that the disk
requires a central hole in order to generate the apparent minimum
of stellar emission that occurs between P1 and P2.

The cluster of blue stars occurs at still smaller
radii, $r<0\asec1,$ and must be closely bound to the black hole,
which was first inferred to exist
from ground-based spectroscopy \citep{d84, dr, k88, r90}.
\citet{kb} carefully registered their high spatial-resolution
spectroscopy to the WFPC2 imagery of \citet{l98}, concluding that the
M31 black hole was coincident with the cluster and that both were
within $0\asec07$ of the M31 bulge photo-center.
The conclusion that the cluster hosts the black hole
was further confirmed by \citet{bend},
who used STIS spectroscopy to show that the central velocity
dispersion in the cluster rises to $1183\pm200~\rm{km~s^{-1}},$
with organized rotation peaking at $618\pm81~\rm{km~s^{-1}}.$
Building on the analysis of \citet{kb}, they reduced the
offset of the black hole from the bulge center to $0\asec033$
in the ``anti-P1'' direction along the P1-P2 line.
They further demonstrated that this slight offset is balanced
by the asymmetry of the Tremaine disk --- the center of mass
of the complete nuclear system is coincident
with the bulge photo-center to $<0\asec01.$

\citet{bend} also designated the cluster as a third component, P3,
of the inner nucleus.
This designation includes a subtle redefinition of P2 as it previously had
been introduced by \citet{l93} and used by \citet{l98}.
The center of the cluster is coincident with the peak of a shallow
surface-brightness cusp seen in V-band that falls within $0\asec1$
of the bulge center;
it was this location that the two Lauer et al.\ papers adopted
as the center of P2.
\citet{bend}, however, use P2 to denote the elongated
and more diffuse concentration of older stars that extends
from the cluster on the side opposite from P1.  In this schema,
P1 and P2 explicitly correspond to the apo- and peri-center apses of the
Tremaine disk, which has a mean peri-center that is {\it not} coincident
with the cluster.

The present environment of P3 appears to be largely quiescent,
despite the immediate proximity of the black hole
\citep[see the extensive discussion of this topic by][]{li}.
\citet{crane} find a weak radio source coincident with P3,
and \citet{garcia, garcia2} find low-level X-ray emission associated with P3,
as well \citep[however, see][]{li}.
On the other hand,
a high spatial-resolution map of emission from ionized gas within in the
nucleus finds a few sources within an arcsecond of P1,
but nothing coincident with P2 or P3 \citep{delb}.
While central accretion of gas may have created the P3 stars in the past,
there is no evidence that any significant reservoir of cold
gas exists there at present.

\subsection{A Closer Look\dots}

Under the \citet{bend} description of P3, the cluster is a disk comprising only
a few hundred A-stars.  Given the scale length of the disk, we concluded that
it should begin to show resolution into discrete sources in images with
slightly higher angular resolution than were obtained by \citet{l98},
such as could be provided by the HST ACS/HRC in the blue.
The photometry of the brighter sources might be obtained directly,
but in any case, the full image of the cluster could also be analyzed by
surface-brightness fluctuation (SBF)
analysis \citep{sbf} to constrain the P3 stellar population.
If the cluster indeed
comprises main sequence A stars, it would look ``lumpy'' in HRC images.
SBF could quantify the amplitude of the lumpiness, and in turn the typical
luminosity of the stars that the cluster comprises.

\section{Obtaining the Highest Spatial Resolution Image of the Nucleus}

We obtained images of the M31 nucleus under GO-program 10572,
using the HST ACS High Resolution Channel (HRC), which provided the best
angular resolution available on the telescope.
Images were obtained in filters F330W ($\rm{U_{330}}$)\footnote{We have
designated the F330W filter band-pass as `$\rm{U_{330}}$' to avoid confusing
it with the standard Johnson U-filter,
which has a significantly different passband.
All magnitudes are on the Vega system, expect as noted.}
and F435W (B) to best enhance the visibility of the blue cluster against the
contrast of the older, redder population of the larger surrounding nucleus.
Even the fine scale of the HRC ($0\asec028\times0\asec025$) undersamples the
point-spread function (PSF) in the blue, however;
we thus dithered the exposures in a $2\times2$
square pattern of 0.5 pixel sub-steps (respecting the different
angular sampling-pitches of the two CCD axes).
In detail, the B-band images were obtained in a
single orbit comprising two dither-sets with a total of eight 298s exposures.
The two sets were offset by 3.5 pixels to reject hot-pixels and other
fixed defects.  The reduced flux and sensitivity of the camera in the
$\rm{U_{330}}$-band required three orbits to obtain an adequate signal.
A complete dither set was obtained in each orbit for a total of
12 672s-exposures.  The three sets were offset from each other in an L-shaped
pattern of $0\asec125$ steps, again to reject CCD defects.
The total peak signal levels per pixel are $\sim 3100~e^\_$ in P3
and $\sim 1600~e^\_$ in P1 for the $\rm{U_{330}}$ image;
and $\sim 8400~e^\_$ in P3 and $\sim 7600~e^\_$ in P1 for the B image.

The goal of the image reduction was to produce a single summed-image
for each filter with a $2\times$ finer scale,
sufficient to ensure Nyquist sampling.
Because the images were obtained with non-redundant 
pointings, the image construction was done in stages
to eliminate cosmic ray events and other defects.
The sub-pixel dither-patterns were executed very
precisely, thus each dither set could be simply interlaced to produce
an initial up-sampled image at the given pointing, with the images
within the set used to recognize and repair (with interpolation)
cosmic-ray events by comparing any one image to its three neighbors.
The two B and three $\rm{U_{330}}$ interlaced images were then
combined into a single image for each filter.
This initial estimate of a Nyquist-sampled image
was then used in a repeat of the first reduction cycle to
go back to and improve the initial detection and repair of cosmic ray events
(and fixed CCD defects) in the source images.

With the second cycle of defect repair completed, the source images
could be then combined more precisely to make a Nyquist-sampled image
directly, using the Fourier-method of \citet{l99}.  This procedure
combines the images in the Fourier domain to algebraically remove
the aliased power present in the source images; significantly, it
handles departures from the optimal dither pattern and combines the
complete data set into the optimal average
Nyquist-sampled image (in a least-squares sense).
Crucially, the \citet{l99} method of image reconstruction is used
in preference to {\it Drizzle} \citep{driz}.
The Fourier-method requires no parametric choices and avoids
the blurring inherent in {\it Drizzle} reconstructions.

\begin{figure}
\plotone{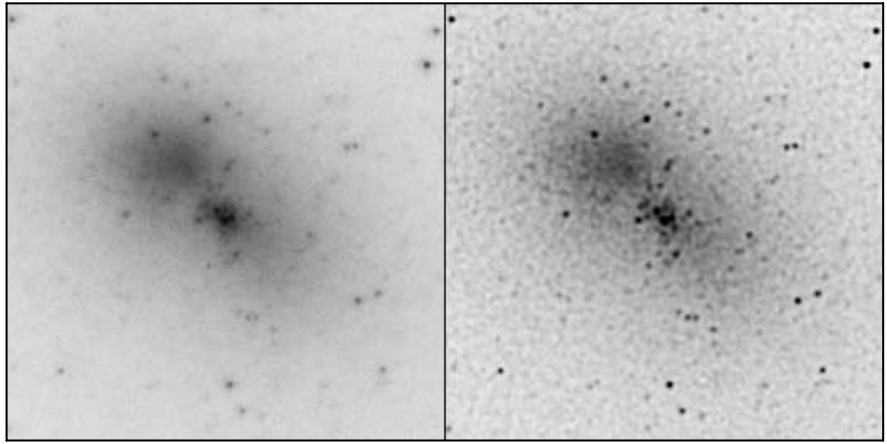}
\caption{The central $3\asec00\times3\asec00$ of the
$\rm{U_{330}}$ (F330W) ACS/HRC image
of the M31 nucleus is shown before (left) and after (right) deconvolution.
North is at the top, and east to the left.
The P3 cluster is at the center of the two panels.
P1 is the diffuse concentration of starlight towards the NE, while P2
is the less prominent extension of light to the SW.
Note that the area shown is still within the overall nuclear cluster.
The M31 bulge only dominates at still larger radii.}
\label{fig:u_fig}
\end{figure}

\begin{figure}
\plotone{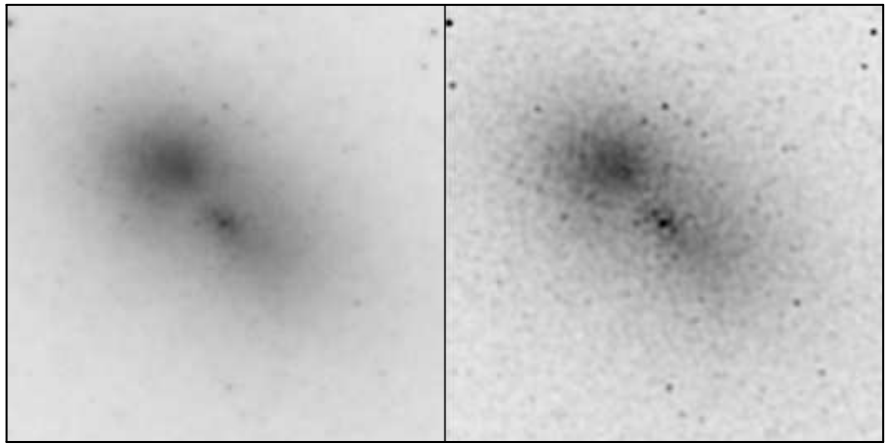}
\caption{As in Figure \ref{fig:u_fig}, but for the B (F435W) ACS/HRC image
instead.}
\label{fig:b_fig}
\end{figure}

The final reduction step for the $\rm{U_{330}}$ and B images was to reproject
them to rectilinear sampling, removing the inherent angular shear
and the rectangular $0\asec028\times0\asec025$ pixels present
in the native HRC image-sampling.  This was done with reference to the
the STScI field-distortion tables for the HRC, which vary with filter.
Sinc-function interpolation was used to compute the undistorted images.
This is the only appropriate interpolation algorithm to use for
Nyquist-sampled images; it does not degrade the resolution in any way.
The final pixel scale of the images was selected to be $0\asec0114,$ which
is 1/4 of the WFPC2/PC pixel scale, to allow for convenient comparison of the
present HRC images to the V and I M31 nucleus images presented in \citet{l98}.
The Nyquist-sampled, rectified, and summed $\rm{U_{330}}$
and B images of the M31 nucleus
are presented in Figures \ref{fig:u_fig} and \ref{fig:b_fig}.

The spatial resolution of the $\rm{U_{330}}$
images is $0\asec039$ FWHM, or 0.15 pc at M31,
putting this among the highest resolution images obtained by {\it HST.}
As sharp as this PSF width is, however, it is
actually significantly broader than a purely diffraction-limited
PSF at 330 nm, which has $0\asec029$  FWHM.
This is due in part to the blurring associated with the
HRC pixel-kernel; however, PSF deconvolution can largely
correct for this, as well removing the wings of the PSF,
yielding nearly diffraction-limited images at both $\rm{U_{330}}$ and B.
Figures \ref{fig:u_fig} and \ref{fig:b_fig} also show
the nuclear images after
40 iterations of Lucy-Richardson \citep{lucy, rich} deconvolution.
The PSFs required were generated
from bright stars present in the outskirts of the images.
Tests presented in \citet{l98} show that this deconvolution procedure
is well-suited to recovering the intrinsic light distribution of
the M31 nucleus.  The final FWHM after deconvolution is $0\asec030,$
which corresponds to a physical resolution at M31 of 0.11 pc.

\section{The Structure and Stellar Population of the Blue Cluster}

The UV-bright P3 star-cluster clearly dominates the $\rm{U_{330}}$ image.
The cluster has a clumpy appearance in this filter,
and further breaks up into apparently discrete sources
in its outskirts (about $>0\asec05$ from the center of the cluster).
P3 is still bright in the B image; however, it appears to be less
clumpy and more concentrated.  As we show below, the source
that appears to define the center of P3
(S11 in Table \ref{tab:stars}) is somewhat redder than
the surrounding cluster, while the discrete sources in the cluster
outskirts are especially blue.  The center of P3 thus increases
in relative prominence going from the $\rm{U_{330}}$ to the B band-pass.
The center of P3 also remains unresolved.  S11, and
a secondary maximum (S07 in Table \ref{tab:stars}) only offset from
it by $0\asec05,$ are both extended rather than stellar.
Both sources likely comprise the blended light of several bright stars.

At large radii, both the $\rm{U_{330}}$ and B images show
the presence of a number of hot stars distributed throughout
the nucleus.\footnote{As noted above, the nucleus dominates the
center of M32 for $r<3''.$ All the images shown in this paper
are limited to radii interior to the nucleus.}
This is the UV-bright stellar population observed by
\citet{king} and \citet{brown}, although
these authors did not characterize sources within $1\asec5$ of the nucleus.
The visual impression is that the hot stars outside of P3 appear to
be weakly concentrated about the cluster, but show no obvious correlation
with the P1 or P2 concentrations of the Tremaine disk.
 
\begin{figure}
\plotone{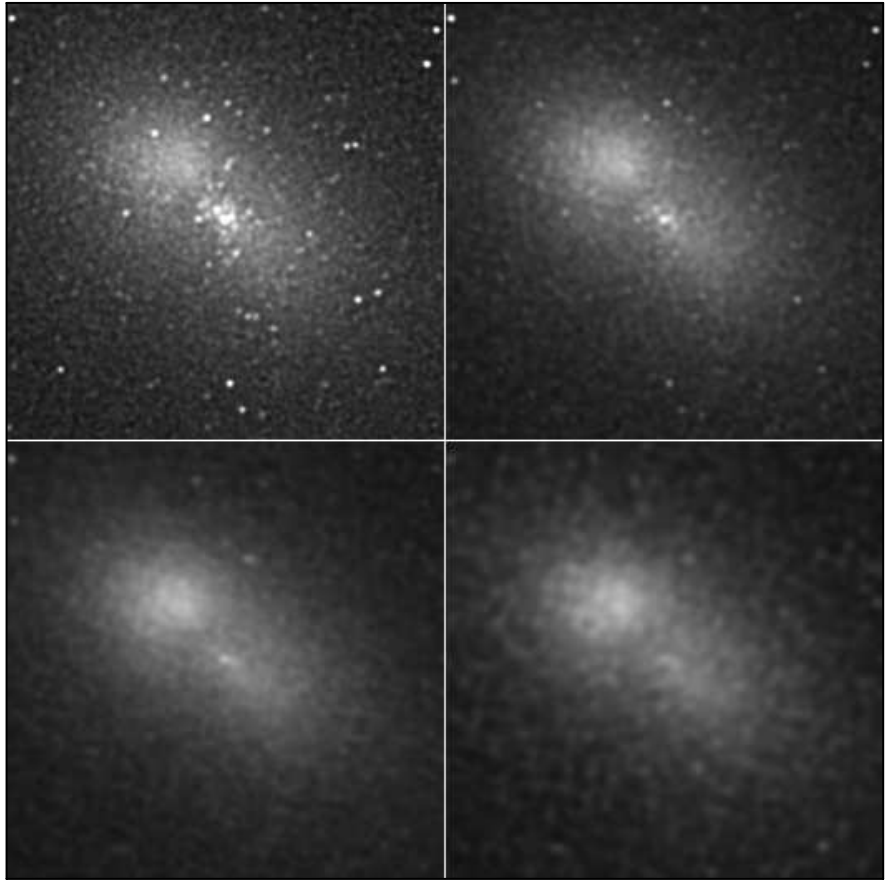}
\caption{The prominence of the P3 cluster of blue stars as a function of
wavelength is evident in this montage showing images of the M31
nucleus in four different bands.
The upper left panel is the $\rm{U_{330}}$ HRC image,
the upper right panel is the B HRC image, while the lower left and right
panels are the V and I band WFPC2/PC images presented in \citet{l98}.
All images are deconvolved and normalized to have the the same
relative surface brightness at the center of the diffuse P1 component.
While the blue cluster dominates the image in $\rm{U_{330}},$ it is essentially
invisible at I.  Each panel is $3\asec00\times3\asec00.$
North is at the top, and east to the left.}
\label{fig:ubvi_fig}
\end{figure}

\begin{figure}
\plotone{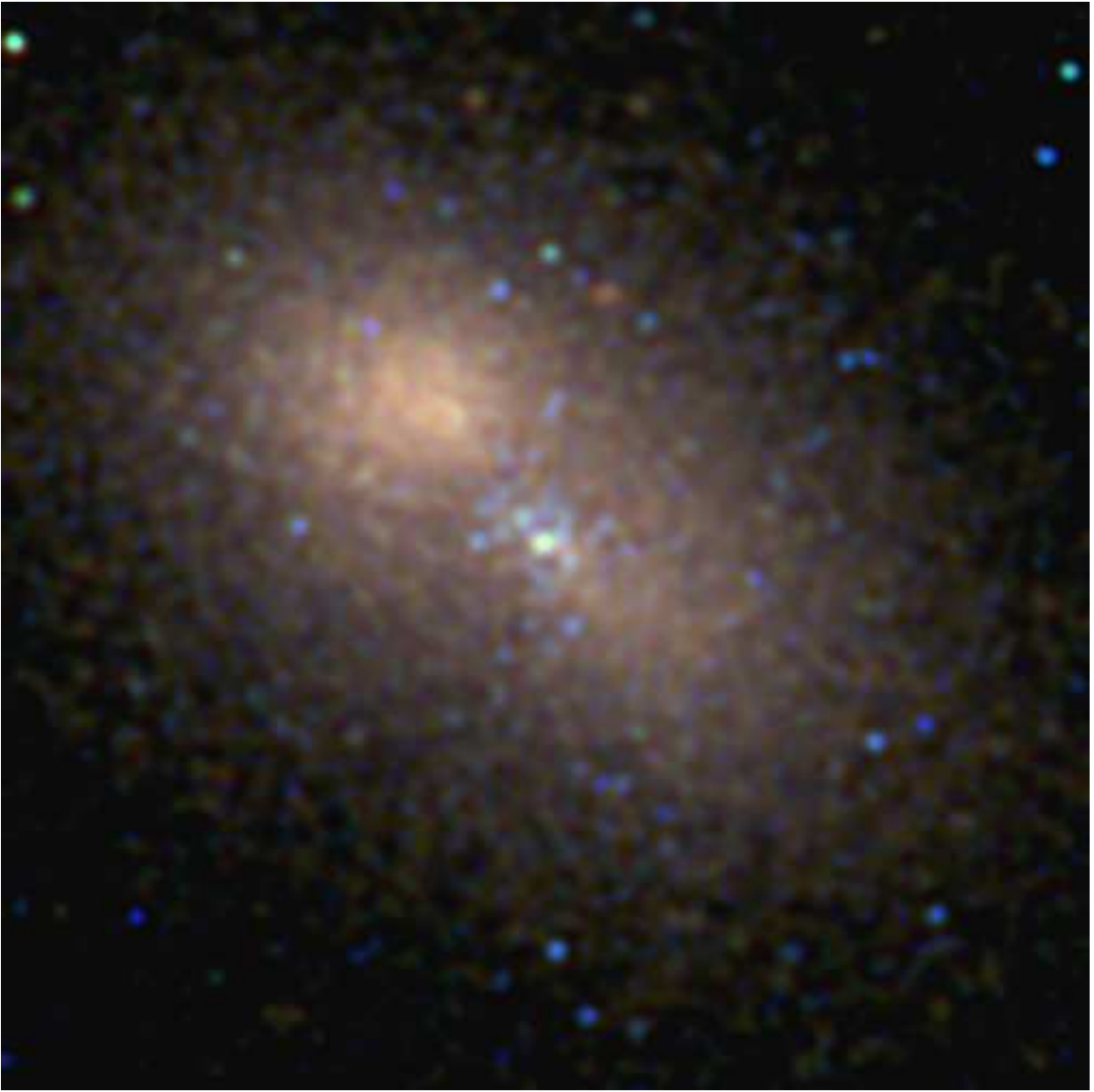}
\caption{The V, B, and $\rm{U_{330}}$-band deconvolved
images of the M31 nucleus are combined as an RGB triplet
in an attempt to generate a plausible ``real color'' image of the nucleus.
The scale and orientation are as in Figures \ref{fig:u_fig}-\ref{fig:ubvi_fig}.
The graininess of the image is due to the SBF pattern intrinsic
to the nucleus.}
\label{fig:color}
\end{figure}

P3 is much less prominent at redder band-passes.
It is essentially invisible at I-band, and shows up only weakly
in V-band, as can be seen in Figure \ref{fig:ubvi_fig}, which compares
the present HRC images to the WFPC2/PC images presented in \citet{l98}.
For this comparison, all four images were normalized to have the same
surface brightness at the center of the diffuse P1 component.
The WFPC2/PC images had already been subsampled to a scale $2\times$
finer than that of the native PC-scale, and were readily shifted, interpolated,
and rotated using sinc-function interpolation to a final
scale $4\times$ finer than the native PC scale.  Figure \ref{fig:color}
shows the $\rm{U_{330}},$ B, and V images combined to make a
representative color image.  The reddish color of the Tremaine disk
contrasts strongly with the hot blue P3 population.  In passing, we note
that the grainy texture of P1, P2, and the surrounding nucleus in
the deconvolved images is due to strong intrinsic SBF.
Close comparison of the B and V images
show them to have correlated SBF patterns, an impressive agreement, given
that the images were taken with different HST cameras, with differing
PSFs and pixel scales.  At $0\asec03$ angular resolution the image ``noise''
characteristics are dominated by the finite number of stars
making up the nuclear components, not the number of photons collected
by the instruments.

The reduced visibility of P3 at V and I highlights the
distinction between it and P2, under the revised schema of \citet{bend}.
P2, which would now be
considered to be the stellar emission associated with the {\it peri-center}
portion of the Tremaine disk, is visible in the V and I images as a
highly elongated feature extending beyond P3 in the ``anti-P1'' direction.
The centroid of this emission is clearly displaced from P3.
Figures \ref{fig:u_fig} and \ref{fig:b_fig} are shown as negative
prints, while Figure \ref{fig:ubvi_fig} is given a positive gray-scale to
offer some diversity in representing the inner nucleus.  The negative
gray scale in the first two figures, for example, appears to accentuate
the diffuse toroidal morphology of the Tremaine disk.

\subsection{Resolved Sources in the Central 1pc of P3}

\subsubsection{Isolation of P3 from the Surrounding Nucleus}

To understand the properties of the P3 system, we attempted to isolate it
from the surrounding nucleus.
Since the cluster is much less prominent
at V and B or $\rm{U_{330}}$, we used the V image as a proxy for the
``background'' light distribution for both the B and $\rm{U_{330}}$ images.
The correct scaling required to subtract the background
was somewhat problematic, given
variations in color among the different nuclear stellar populations.
As shown in \cite{l98}, P1 is significantly redder than the
outer nucleus, having $\Delta V-I=0.07,$
with the portion of P2 outside P3 in the
``anti-P1'' direction having an intermediate color.  A simple hypothesis
is that the Tremaine disk, which comprises both P1 and P2, is
redder than the outer portions of the nucleus.  P2 appears to be
slightly bluer than P1, since it has lower contrast against the
underlying nucleus.

The redder color of the Tremaine disk as compared to the surrounding nucleus
means that there is no simple scaling of the V to the $\rm{U_{330}}$ or
B images that will yield a completely flat background around P3.
Scaling the V image to the outer nucleus in the $\rm{U_{330}}$
image over-subtracts the Tremaine disk,
leaving large negative residuals at P1, as well as somewhat reduced but still
negative residuals in the P2 region immediately surrounding P3.
Conversely, scaling the two images by their P1 fluxes, leaves P3
in a significantly positive background.  Our solution was to scale
to the portion of P2 immediately outside the apparent extent of the
P3 cluster in the anti-P1 direction.  This still yields a positive
background in the outer nucleus, and negative residuals in P1, but does
produce a low background surrounding P3.  We also ``clipped out''
residual portion of P3 still present at V, and slightly smoothed the
image as well, to avoid biassing the measurement of the central
fluxes of P3 in $\rm{U_{330}}$ or B,
or introducing artificial structure to either image.

\subsubsection{Stellar Photometry of the P3 Sources}

\begin{figure}
\plotone{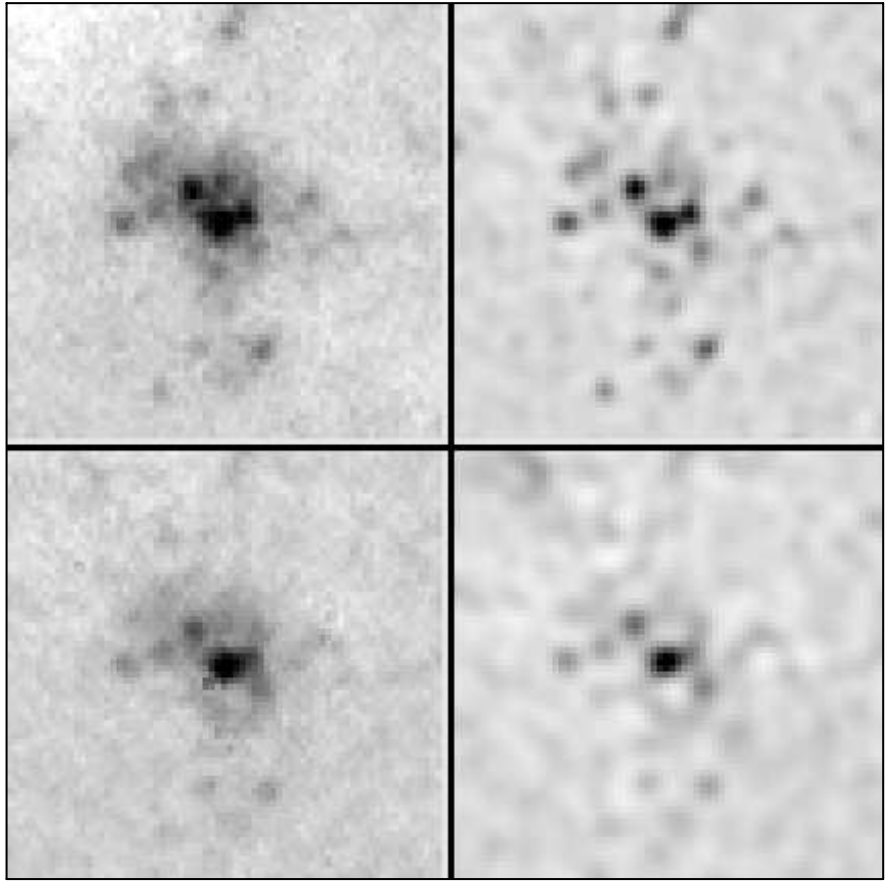}
\caption{The P3 cluster is shown for $\rm{U_{330}}$
(top row) and B (bottom row) with
the background nucleus and bulge subtracted.  Images are from
the reduced Nyquist-sampled images (left column) and after
deconvolution (right column). The pixel scale is $0\asec0114.$  Each
image is $71 \times 71$ pixels, which corresponds to $0\asec81\times0\asec81$
or $3\times3$ pc at M31.  North is at the top,
and east to the left.}
\label{fig:p3}
\end{figure}

The isolated P3 cluster before and after deconvolution in
the $\rm{U_{330}}$ and B filters is shown in Figure \ref{fig:p3}.
This figure zooms into the cluster, ratifying the visual impression
stated at the start of this section that the outskirts of the
cluster breakup into individual point sources.  The deconvolved images
show little diffuse emission outside the very center of the cluster.
Close comparison of the $\rm{U_{330}}$ and B images shows that the
sources cover a range in color.  There are a number of sources
that are readily visible in $\rm{U_{330}},$ but nearly invisible in B
(which unfortunately give rise to large errors in the
${\rm U-B}$ values of the sources).
The bright source, S11, at the center of P3 is clearly extended,
as is S7, the source adjacent to it;
however, to the degree of resolution
afforded by these observations, most of the other sources appear to be stellar.

Measuring the fluxes of the P3 sources is difficult with traditional
PSF-fitting techniques,
given the high degree of crowding and the strongly variable background.
Instead, we used a novel procedure that measures stellar fluxes
from aperture photometry on deconvolved images,
which has been shown to work extremely
well on Nyquist-sampled ACS/HRC images of M32 \citep{m32}.
The methodology is to take the deconvolution of the $\rm{U_{330}}$
and B frames to 640 Lucy-Richardson iterations.
This effectively transforms the observed PSF into a highly compact
``deconvolved PSF'' that
essentially concentrates all of its flux to within a radius of $0\asec02.$
Aperture photometry is done on the sources in the resulting image,
once the background nucleus light is subtracted off.
The resulting colors and magnitudes of the sources within 1 pc of P3 are
mapped in Figure \ref{fig:stars} and listed in Table \ref{tab:stars}.
A comparison of this figure to the images in Figure \ref{fig:p3},
ratifies the impression that the bluer sources occur on the
outskirts of the P3 cluster.

Interpreting the present photometry is complicated by the
large color-terms required to bring the F330W into concordance
with the standard Johnson U band-pass.  Based on the color equations
presented by \citet{acs}, for Johnson ${\rm U-B<0.2}$ we derive
the transformation from the F330W to U bandpass as
\begin{equation}
\rm{U-U_{330}=0.116-0.406(U-B)+0.125(U-B)^2.}
\end{equation}
Fortunately, the difference between the HRC F435W filter and Johnson B
only varies by a few hundreds of a magnitude over the large F330W-F435W
color range of the observations, and can be directly taken as B, given
the larger random errors in the photometry.
The analogous transformation for ${\rm U-B>0.2}$ is
\begin{equation}
\rm{U-U_{330}=0.160-0.104(U-B)-0.093(U-B)^2.}
\end{equation}
As can be seen from the equations above, and the transformed values
in Table \ref{tab:stars}, the implied corrections are as large
as $\sim40\%$ of the observed $F330W-F435W$ colors.  While
\citet{acs} present calibrating observations over the needed
range of ${\rm U-B,}$ the actual corrections will depend on the unknown
metalicities and atmospheric properties of the actual P3 stars.

\begin{figure}
\plotone{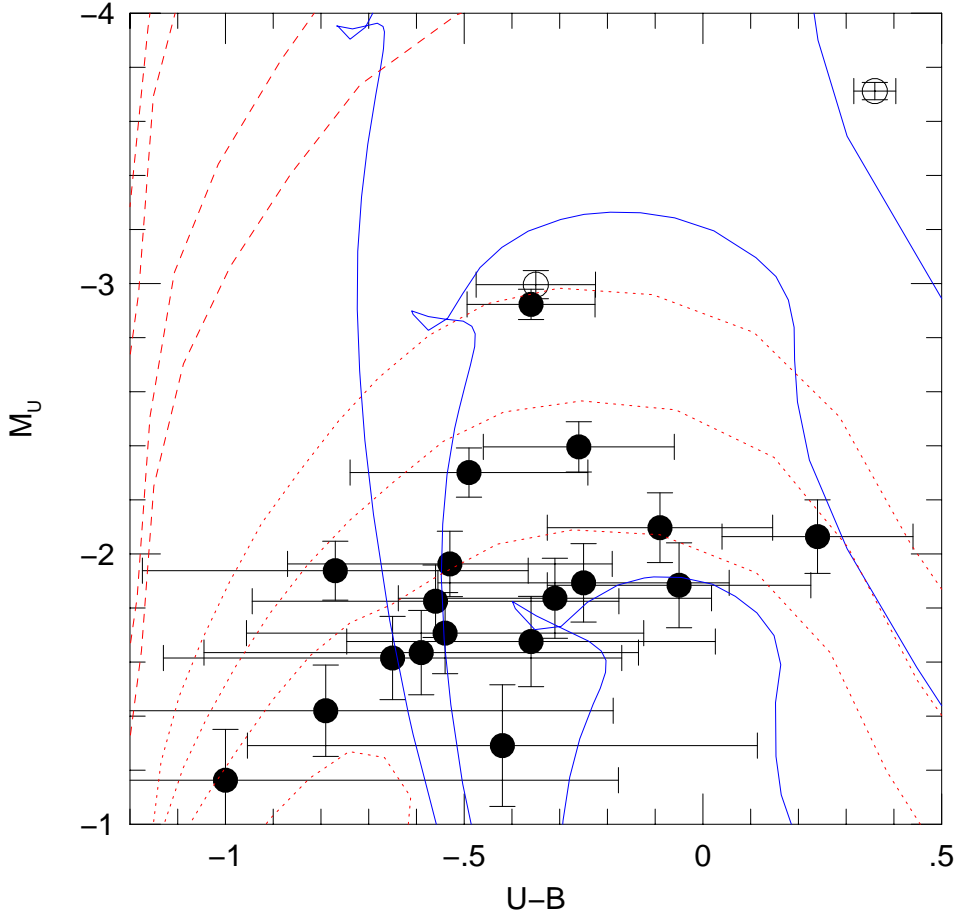}
\caption{A $\rm{U-B}$ vs. U color magnitude diagram is shown for the
sources within 1pc of the P3 center.  The open symbols denote S7, and S11,
which are clearly spatially extended.  S11, the nominal center of P3,
is the brightest and reddest object plotted.
Padova isochrones \citep{pad} (blue lines)
are shown for single-burst populations
with ages of 50, 100, and 200 Myr going from the top down.
Contrasting PAGB (red dashed-lines) and PEAGB (red dotted-lines)
tracks (Bressan et al., in prep.)
are shown to test the hypothesis that some of the P3 stars
may really be old stars in late evolutionary phases.
Going from bottom to top, are one  AGB-manqu\'e and
three PEAGB tracks with core
masses of 0.50, 0.54, 0.56, and 0.58 $M_\odot;$ and four
PAGB tracks are for core masses of 0.57, 0.60, 0.75, and 0.90 $M_\odot.$}
\label{fig:cmd}
\end{figure}

We plot the colors and absolute magnitudes of the P3 sources
in Figure \ref{fig:cmd}.
We attempted to correct for both foreground and M31 internal dust absorption.
The galactic foreground extinction provided by \citet{sfd}
is $A_V=0.206$ mag, which given the reddening tables in
\citet{acs} yields $A_B=0.26$ and $A_{U_{330}}=0.31$ mag.
A survey of the literature shows that internal dust extinction
in the M31 bulge is small \citep[e.g.][]{ttt}.
Hui Dong (private communication) estimates internal $A_V=0.08$
for the region of the bulge just outside the nucleus based
on stellar-population models of multi-color {\it HST} imagery.
The implied total values are then $A_B=0.36$ and $A_{U_{330}}=0.44$ mag.

Figure \ref{fig:cmd} plots Padova \citep{pad} isochrones (blue)
for single-burst stellar populations of 50, 100, and 200 Myr.
It appears that most, if not all, of the sources are consistent
with an age of $\sim100-200$ Myr.  We note that the brightest
source in the cluster, S11 in Table \ref{tab:stars},
is also the reddest source; it is extended as well.  S11 is at the
center of P3, and we have adopted it is the {\it ad hoc} center of the
system for the coordinates given in Table \ref{tab:stars}
and Figure \ref{fig:stars}.

At the same time, evolutionary tracks for some late evolutionary-phases
of old stellar populations pass through the U and ${\rm U-B}$
domain occupied by the P3 stars, as well.
These are all due to metal-rich giants that have evolved
into ``extreme horizontal branch'' (EHB) stars and later move
back to the red as their core helium-burning phase is completed.
Stars that populate the EHB have lost a significant amount of their
envelope on the RGB, though the mechanism causing the mass loss 
is not precisely known \citep[for a full review see][]{greg,ocon}.
These stars do not have a large enough convective envelope to allow a
full AGB phase. Instead, the hottest of these EHB stars will fail to
reach the AGB track at all, and evolve through the AGB-manqu\'e
channel. Cooler EHB stars will eventually leave the AGB track early
and become Post Early-AGB stars (PEAGB).
These phases, collectively called hot post-horizontal branch (HP-HB) evolution,
have been previously found in M31 \citep[e.g.,][]{king,brown,bert}.
Their surface density has recently been shown to increase towards
the center of M31 (Rosenfield et al., submitted) more rapidly
than the underlying bulge surface brightness,
although there appears to be no enhancement associated with the
overall nucleus, itself.

Figure \ref{fig:cmd} shows several examples of HP-HB tracks,
such as H-burning Post-AGB (PAGB)
tracks of masses $0.57 < M_\odot < 0.90$ (red, dashed)
from \citet{vass}  with $Z=0.016$ and $Y=0.25,$
and AGB-manqu\'e and PEAGB tracks
(red, dotted) of masses $0.50 < M_\odot <0.58.$
These are from the most recent
Padova stellar evolution library (Bressan et al., in prep.)  with $Z =
0.07$ and $Y = 0.389$ and an $\alpha$-enhanced composition typical of
elliptical galaxies \citep[adapted from][]{bens}.
Tracks were converted to Johnson filters following the color transformations of
\citet{pad}. The metallicity of the Padova tracks was chosen as
an extreme case to allow for high helium content, ensuring high mass
loss on the RGB. As these tracks are preliminary, other metallicity
and mass loss rate combinations could also produce HP-HB stars with
similar luminosities and effective temperatures.

Figure \ref{fig:cmd} shows that the canonical
PAGB tracks are too bright and too hot to explain the P3 sources;
however, a number of the P3 sources, especially the bluest and faintest ones,
are consistent with the PEAGB evolutionary tracks shown
(and the brightest of the AGB-manqu\'e tracks, which is the faintest track
shown).
Despite this concordance, however, we suggest that only a few
of the P3 stars could be in the HP-HB phase, in contrast to
being hot newly-formed stars.   HP-HB evolution across the U and ${\rm U-B}$
domain in Figure \ref{fig:cmd} is very rapid, occupying only $\sim10^5$ yr.
The P3 stars are highly concentrated, suggestive of
a local population tightly bound to the black hole,
but there is no trace of the associated substantial population of giants
required to generate and maintain a short-lived UV-phase
sufficient to account for the complete UV luminosity of P3.

\subsection{Surface Brightness Fluctuation Analysis of the Cluster Population}
\label{sec:sbf}

A separate approach to characterizing the stellar population of P3
is to use ``surface brightness fluctuations" (SBF)
as a diagnostic \citep{sbf}.  In this case we are effectively using the
``clumpiness" of P3 as an indicator of the typical luminosity
of the stars that characterize the variance in surface brightness.
For a constant total luminosity
of the P3 cluster, the variance about a perfectly smooth distribution
will steadily increase as we go from modelling it as the aggregate
of a large number of intrinsically faint old stars to a small handful
of high luminosity young stars.

\begin{figure}
\plotone{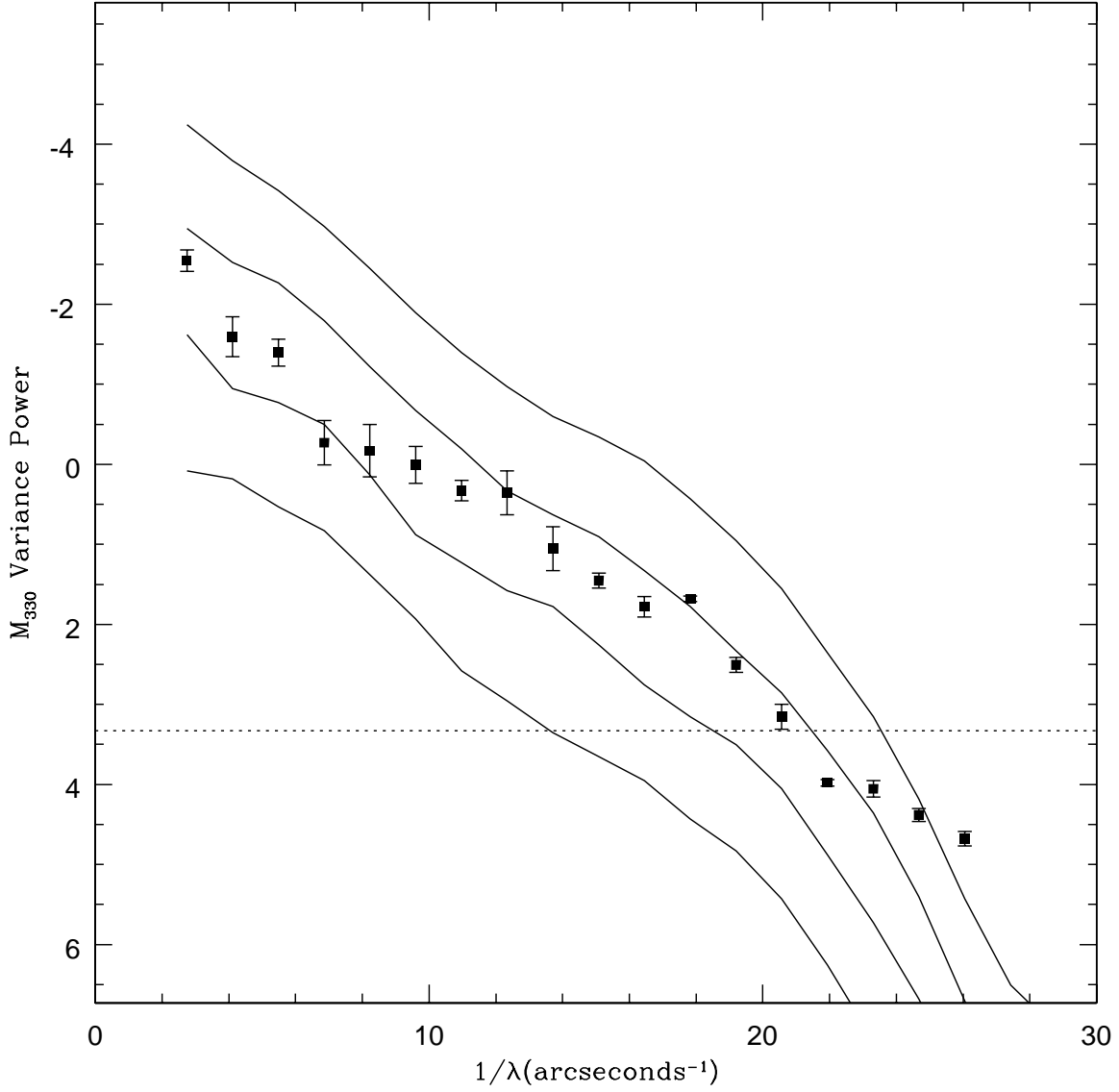}
\caption{The SBF power spectrum for the P3 cluster in the ${\rm U_{330}}$
image is shown (square symbols), compared the to the average power spectra
for models constructed from 50, 100, 200, and 400 Myr single-burst
stellar populations (lines from top to bottom).  The dotted horizontal
line shows the noise level in the ${\rm U_{330}}$ image.  The ordinate
gives the power in terms of absolute ${\rm U_{330}}$ luminosity
corrected for extinction.  The top of the scale corresponds to the
${\rm U_{330}}$ absolute magnitude of P3.}
\label{fig:sbf}
\end{figure}

The procedure for conducting an SBF analysis on P3 requires measuring
the power spectrum from the images of the cluster, and comparing it
to the power spectra of models of the cluster.
Measurement of the observed power spectrum begins with
subtracting a smooth model of the P3 light distribution
(see $\S\ref{sec:sb}$), and then measuring the power spectrum of the
residual image.  Since the SBF technique uses the shape of the
PSF power-spectrum as a template, the analysis is done on the
background-subtracted image of P3 in the observed, as opposed to the
deconvolved domain.  In detail, we restricted the analysis to a
$0\asec5$ radius aperture centered on P3.  We also only performed
the SBF analysis on the ${\rm U_{330}}$ image, given concerns about the
accuracy of the technique in the B image, which has a significantly
reduced contribution from P3, but increased background variance (which
cannot be subtracted) from the underlying old-population of the nucleus.
The ``raw'' power spectrum of P3 includes a constant noise term that
largely represents random photon shot-noise in the image. We
subtracted this off by measuring the power a spatial frequencies
higher than the PSF Nyquist scale.  We show the resulting power
spectrum in Figure \ref{fig:sbf}.

The simulated power spectra comes from
models constructed for bursts of 50, 100, 200, and 400 Myr,
drawing stars at random from the Padova \citep{pad} isochrones
transformed to ${\rm U_{330}}.$
The location of the stars was drawn at random from the smooth
model of the P3 light distribution until the total ${\rm U_{330}}$
luminosity of P3 was reached.
The models were then analyzed by the same procedure
used for the real data.  A few dozen models were generated for each burst
age to average out statistical fluctuations in the power spectra.
The power spectra in Figure \ref{fig:sbf} are plotted in terms
of the implied absolute ${\rm U_{330}}$ magnitude of the flux.
The intercept of the curves at zero spatial-frequency gives the
luminosity of the \citet{sbf} $L_*$ parameter, which characterizes
the variance as due stars of that representative brightness.
The limit of the vertical scale is set by the luminosity of P3, itself.

The observed SBF power appears to
fall between that of the 100 and 200 Myr bursts,
which is qualitatively
ratified by the subjective appearance of the P3 models of the same age.
This result is thus in excellent agreement with the P3 CMD
analysis presented in the previous section.

While we can constrain
the age of the P3 population, it appears, however, that we have little
sensitivity to the form of the initial mass function (IMF), given
the limited dynamic range of the luminosity of the sources seen in P3.
\citep{bart} argue that the stars surrounding the Milky Way black hole have
a markedly ``top-heavy'' IMF.  We generated 200 Myr SBF models with
their $dN/dm\propto m^{-0.5}$ IMF to see if this might produce
significantly different SBF power as compared to that for the same age derived
from the Padova isochrones.
We only measured a modest increase
in SBF power, however, which would cause us to infer a slightly older
age for the population by $\sim50$ Myr (it takes an older ``top-heavy''
model to match the observed SBF power of P3, itself),
a result that is within the error limits of the SBF results already quoted.
The SBF power is dominated by the resolved sources near the main-sequence (MS)
turn-off, which is determined by age alone.  A more top-heavy IMF
qualitatively down-weights the contribution of fainter stars on the MS
to the total SBF power, causing the relative SBF variance to increase
for the same age and total P3 luminosity.
However, given the steep $L\sim m^4$ mass-luminosity
relation for the MS, we have in the end little leverage on the mass function.
A small range in MS mass corresponds to a large change in stellar
luminosity, thus suppressing information on the mass function, itself.
We further note that the CMD analysis has even less sensitivity
to the IMF, given the even more limited mass range of the resolved stars.

\begin{figure}
\plotone{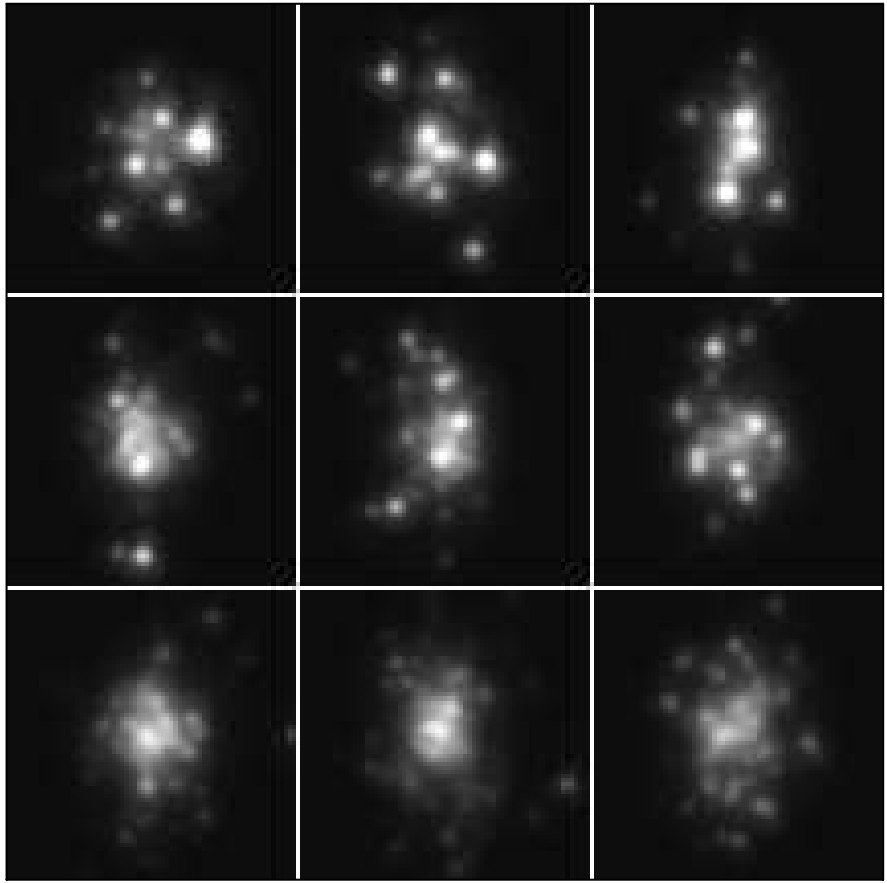}
\caption{Models of the P3 cluster based on the smooth light distribution
derived in $\S\ref{sec:sb}$ and random samples of stars drawn from the
Padova isochrones \citep{pad} for bursts of 50 Myr (top row), 100 Myr (middle),
and 200 Myr (bottom).  Each row shows three different realizations for the
given burst age.  The scale and image size are the same as
in Figure \ref{fig:p3}.}
\label{fig:p3_mod}
\end{figure}

Some sample models for P3 are shown in Figure \ref{fig:p3_mod}.
The clumpiness of the models shown can be seen to steadily decrease
with increasing burst age.  It is notable that the strong SBF
pattern completely dominates the appearance of the simulated clusters.
Further, the location of either the peak brightness,
or light-weighted centroid of the entire cluster can
be displaced significantly away from the
nominal center of the ideal smooth cluster model due
to the large random fluctuations in the structure.
This is an important caveat to respect when evaluating the precise location
of the P3 cluster with respect to the M31 center of mass or
the location of the M31 black hole (see below).  For the ensemble
of 50 Myr models, the average location of the peak flux is
offset from the true cluster center by $0\asec12,$ and the
photo-center is displaced placed by $0\asec05,$ with the
typical scatter from model to model being nearly as large as these
mean values.  For the 100 Myr models, the two offsets decrease
to $0\asec06$ and $0\asec02;$ and $0\asec03$ and $0\asec01$ for the
200 Myr models.

\subsection{The Spectral Energy Distribution of P3}

The total brightness of P3 was measured by aperture photometry
performed on the difference images.  The apparent magnitudes
of P3 are $20.3\pm0.1$ (AB) or $19.1\pm0.1$ (Vega) in $\rm{U_{330}}$, and
$19.0\pm0.1$ (AB) and $19.1\pm0.1$ (Vega) in B, uncorrected
for foreground extinction.  Using the reddening values stated earlier, the
the corrected apparent fluxes are $42\pm4$ $\mu$Jy in $\rm{U_{330}}$ and
$125\pm12$ $\mu$Jy in B.  The implied reddening-corrected colors
are $\rm{U_{330}-B}=-0.1\pm0.2$ and $\rm{B-V}=0.4\pm0.3$ (Vega),
using the $m_V$ measure from \citet{l98}.

\begin{figure}
\plotone{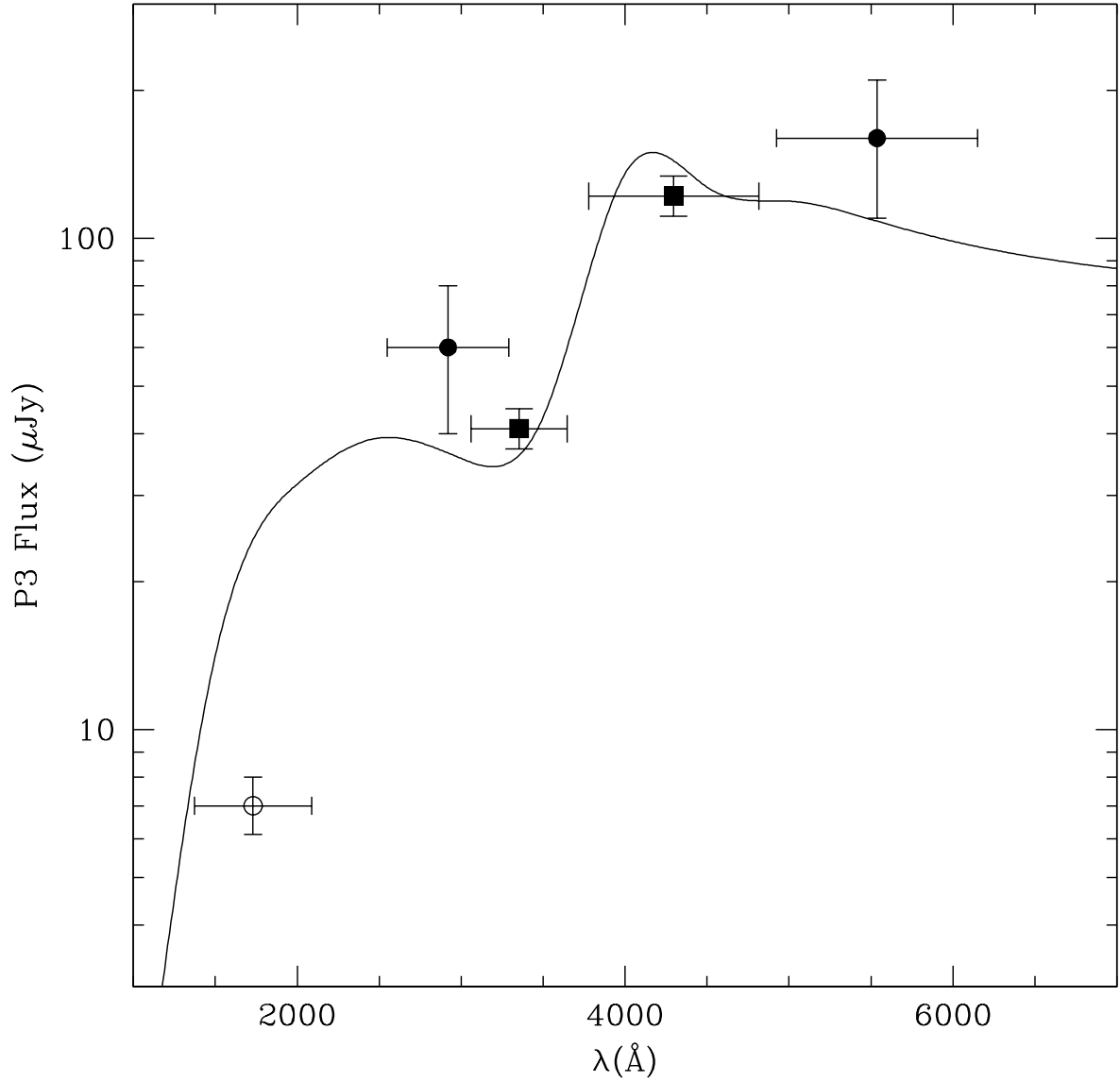}
\caption{The SED of P3 is shown based on the broad-band total magnitudes
of the present study (large squares), \citet{l98} (round dots), and the
FUV flux of \citet{king} (open circle).  The horizontal bars give the
approximate band-widths.  An $f_\nu$ SED template for an A0 V star is shown
scaled to the points for comparison.}
\label{fig:flux}
\end{figure}

The implied spectral energy distribution (SED)
of P3 is plotted in Figure \ref{fig:flux},
based on the FUV 175nm flux measured by \citet{king},
the $\rm{U_{330}}$ and V fluxes
measured by \citet{l98}, the present $\rm{U_{330}}$ and B values.
We reduced the FUV measurement
from $6.7~\mu{\rm Jy}$ to $5.6~\mu{\rm Jy}$,
for consistency with the present reddening corrections.
\citet{king} assumed $A_{175}=0.88,$ based
on the total $E(B-V)=0.11$ for M31 measured by \citet{mr}.
Our $E(B-V)=0.086$ value implies $A_{175}=0.69,$ instead.
An $f_\nu$ SED of an A0 main sequence star scaled
to the flux-points is shown for comparison.
As noted in \citet{l98}, and shown spectroscopically
by \citet{bend}, the SED of P3 is indeed ``A-like.''  The
FUV point does fall well below the template, but this portion of the
SED is highly sensitive to temperature, and the FUV point could be fitted
with a modest adjustment of the template; this point will also be the
most sensitive to the correct value of the extinction.

\subsection{The Surface Brightness Profile of P3}
\label{sec:sb}

\begin{figure}
\plotone{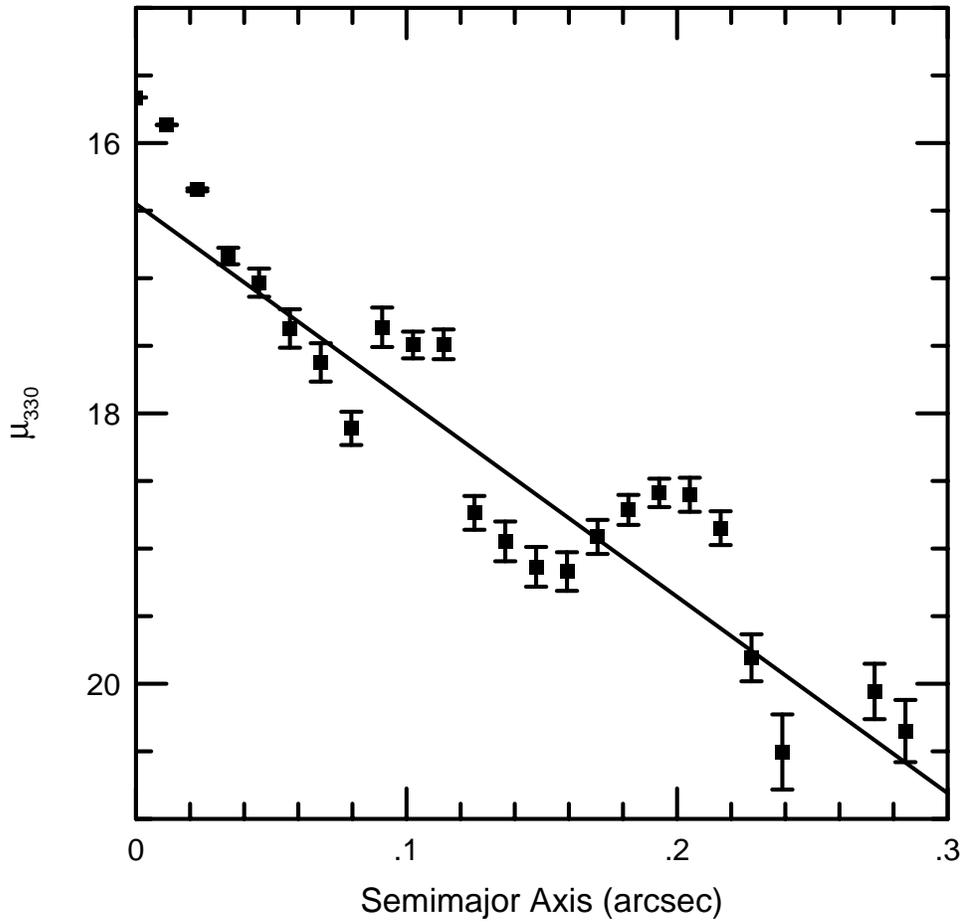}
\caption{The apparent surface brightness of the P3 cluster
in ${\rm U_{330}}$ as a function
of semimajor axis is plotted.  The solid line shows the exponential form
fitted to the profile for $r>0\asec04.$}
\label{fig:p3_sb}
\end{figure}

The over all distribution of light within the P3 cluster is confirmed
to roughly follow the exponential light distribution derived
by \cite{bend}.  A $\rm{U_{330}}$-band surface brightness profile of P3 is shown
in Figure \ref{fig:p3_sb}, with the exponential model shown as a solid line.
S11 was taken as the center of the cluster,
but the fit excludes points with $r<0\asec04,$ given that the inner-most
points will still be affected by the residual PSF associated with the
deconvolution.  The exponential form recovered is
\begin{equation}
\rm{\mu_{330}}(a)=\left(13.4\pm1.7\right)a + 16.45\pm0.32,
\end{equation}
where $a$ is the semimajor axis in arcseconds,
and $\mu_{330}$ is the surface brightness (magnitudes/sq. arcsec)
observed with the F330W Vega-based zeropoint (no reddening
correction has been applied).  The implied exponential scale-length
is $0\asec075\pm0\asec010,$ which is marginally more compact than
the $0\asec1\pm0\asec01$ value derived by \citet{bend}. 
The average ellipticity of the isophotes is $0.33\pm0.03,$ which implies
an inclination of $48^\circ\pm5^\circ,$ if P3 is interpreted to
be a disk, again in good agreement with the $55^\circ\pm2^\circ$
of \citet{bend}.

Although S11 was excluded from the fit, given concerns about the
residual deconvolved-PSF, the brightness profile of S11
actually rises about the exponential model.  The core radius of S11
is $<0\asec024$ or $<0.09$ pc.  If we estimate the volume
luminosity-density of S11 from $\rho_L=I_0/(2r_C),$ where $I_0$ is
the central surface brightness ($<15.53$ in U, corrected for
extinction), then at S11 we find $\rho_L>3\times10^5~L_\odot~
{\rm pc^{-3}}$ in the U-band.

\subsection{The Location of the M31 Black Hole}

The present work provides no new dynamical information about the M31 black hole,
and cannot offer any objective improvement on the conclusion of
\citet{kb} or \cite{bend} that the black hole must be close to the center of P3.
At the same time, we advance subjective arguments that the location of
the black hole may be coincident with S11 (see Table \ref{tab:stars}).
The central location of this source is suggestive.
While we showed in
$\S\ref{sec:sbf}$ that simulated models of the cluster often have
the brightest clump of stars to be significantly offset from the true
center of the cluster (which is where we would expect the black hole to be).
However, while we do not know the ``true'' center of P3,
case, the location of S11 and the photo-center of P3 at least agree
to within $0\asec02.$
S11 is also extended and significantly redder than the other
sources, underscoring its unique properties among the other sources.
Its red color indeed makes it the only source to mark the location of P3 in the
V-band.  A truly extended source should not exist within the strong
tidal field of the black hole, unless it represents the peak of
a cusp made of the stars most closely bound to it. At the same time,
it's possible that the structure of S11 is only due to a random
fluctuation of stars along the line-of-sight.  S07, is also extended,
is offset from the center of P3, but otherwise has no special attributes.
The SBF simulations presented in the previous section show that most
of the sources in P3 are indeed dominated by a single bright star; however,
very close to the center there can be one or two blends of
similarly bright stars within the HRC PSF.

\section{Discussion, Summary, and Conclusion}

We began the paper with a history of earlier work on P3, which already built
a strong case that it is a cluster of young stars.  We have now partially
resolved the cluster into individual stars, and characterized its form with
significantly higher spatial resolution than was provided by the WFPC2/PC
observations of \citet{l98}; however, in broad detail the conclusions
of that paper, and the later work of \citet{bend} remain unaltered.
Both a color-magnitude diagram (CMD) obtained of the resolved stars,
and an SBF analysis performed on the total image,
show that P3 plausibly comprises a population
formed in a burst 100 to 200 Myr ago, consistent with the P3 SED
and the spectrum obtained by \citet{bend}. The result may also be
supported by the \citet{sag} spectroscopy observation of the
inner bulge, which suggest that a burst of star formation occurred
within the nucleus about $\sim100$ Myr ago.

One small modification of the earlier results provided by the present work
is that that P3 may be slightly younger than
the 200 Myr age derived by \citet{bend}, but that work really only ruled
out an age much younger than 100 Myr.  Another new result
is that the color of S11 suggests that a single-age starburst
model for the cluster does not capture the full story of P3.  At some level,
the large luminosity of P3 blends with the much bluer stars around
it to generate an overall A-like SED.  It is intriguing that S11
provides the only visible trace of P3 in the V-band.   If it does mark
the location of the M31 black hole, then it suggests that there
may yet be another change in the stellar population of the nucleus
within $\sim0\asec03$ or $\sim0.1$ pc of the black hole.

Although we have concluded that the P3 stars are young,
\citet{king} and \citet{brown} argue that
the UV-bright stars in the bulge of M31, which of course surrounds
the nucleus, represent the final stages of stellar evolution
in an old metal-rich stellar population.  If there are truly two
different populations contributing UV-bright stars to the bulge and
nucleus of M31, it still begs the question of whether or not
old remnants may be present in P3, as well as how far out the
putative young stars in P3 may be traced out into the surrounding
nucleus or even bulge.  We show that PEAGB tracks traverse
the U vs.\ ${\rm U-B}$ domain defined by the P3 stars, thus it is
possible that a few of the P3 stars are indeed old-evolved stars,
rather than being young massive main sequence stars.
However, since this is a very rapid
evolutionary phase, it is extremely unlikely that the entire
P3 cluster could be accounted for by PEAGB stars.

A strong factor in constraining the population of P3
is the compactness of P3, itself.
The UV-stars outside the nucleus do trace the bulge-light at long wavelengths,
but yet are not strongly concentrated within the outer nucleus; their
abundance by total light or number is but a small faction of the bulge.
Requiring any sort of HP-HB stars as the dominant P3 population
requires an associated massive population of red giants also closely
bound to the M31 black hole to continually generate the short-lived
PEAGB stars.  There is no evidence for such a population
in the V and I nuclear images of \citet{l98}.

This same consideration would appear to also apply to
the hypothesis of \citet{demar}, who suggest that the P3 UV-bright
stars may be the remnant cores of giants stripped by close
encounters with the black hole.
The high velocities associated with the black hole are such that stars
passing close to the black hole from the bulge, or even the outer nucleus
will not dwell in the vicinity of P3.
If the P3 stars are really processed giants, they still must have been
initially closely bound to the black hole,
which again begs the question of where the progenitor giants are.
The only way that this mechanism might work
is if the stripping of the P3 stars takes place just as they
begin their first ascent up the red giant branch.
This might produce long-lived UV-bright stripped-cores,
and would account for lack of progenitor giants as well.

\begin{figure}
\plotone{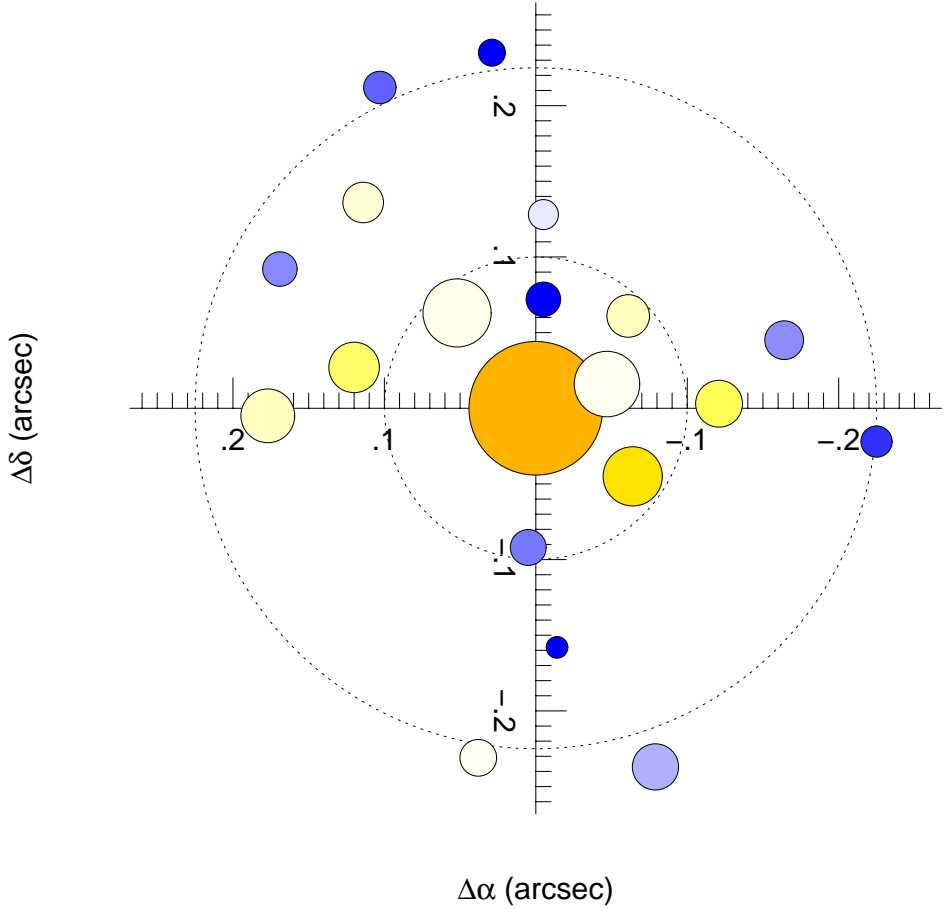}
\caption{A symbolic ``star map'' of the sources in the central parsec
of P3 is shown. The area of the symbols is proportional to
the B band luminosity.  The color encodes the relative ${\rm U-B}$ color
of the sources.
The two central sources, S11 and S7, are extended and are probably blends
of several stars.
The origin is taken as S11 in Table \ref{tab:stars},
which is inferred to mark the location of the black hole.
This map can be directly compared to the images of P3 in
Figure \ref{fig:p3}. The inner and outer circles correspond to orbital
velocities of 1000 and 1500 km/s for stars in circular orbits (in the plane
of the sky)
about a $1.4\times10^8~{\rm M_\odot}$ black hole \citep{bend}.}
\label{fig:stars}
\end{figure}

Further investigation of the origin of P3 might be linked to
understanding the early-type stars closely
bound to the black hole in our own galaxy.  The existence of P3
in a second Local Group galaxy suggests that this is not a rare
phenomenon.  The present observations offer an additional site
to test mechanisms that can form stars within the strong tidal field
of a supermassive black hole.  We finish by noting that as with the
galactic center, it may be possible to measure the proper motions
of stars closely bound to the M31 black hole.  Figure \ref{fig:stars}
shows that circular velocities in the plane of the sky around
the M31 black hole will exceed $1000~{\rm km~s^{-1}}$ over the entire
extent of P3.  It is possible that proper motions may be detected in about
a decade after the present observations were made.

\acknowledgments

We thank Karl Gebhardt, Scott Tremaine, Jessica Lu,
and Pierre Demarque for useful conversations.
We thank Alessandro Bressan and Leo Girardi for advance use of preliminary
Padova stellar evolution tracks.
We thank the referee for a number of useful suggestions.
Pete Merenfeld (NOAO) kindly prepared the three-color
image of the nucleus.

\clearpage

\begin{deluxetable}{rrrccccc}
\tablecolumns{8}
\tablewidth{0pt}
\tablecaption{Sources in the Central Parsec of M31}
\tablehead{\colhead{ }&\colhead{$\Delta\alpha$}
&\colhead{$\Delta\delta$}&\colhead{ }&\colhead{ }&\colhead{ }&\colhead{ }
&\colhead{ }\\
\colhead{N}&\colhead{(arcsec)}
&\colhead{(arcsec)}&\colhead{$m_{330}$}&\colhead{$M_{330}$}&\colhead{$m_{330}-m_{435}$}&\colhead{$M_U$}&\colhead{${\rm U-B}$}}
\startdata
S01&$ -0.225$&$ -0.022$&$22.82\pm0.15$&$-2.05$&$ -1.10\pm0.48$&$-1.61$&$-0.65$\\
S02&$ -0.164$&$  0.045$&$22.54\pm0.12$&$-2.33$&$ -0.93\pm0.34$&$-1.96$&$-0.53$\\
S03&$ -0.121$&$  0.003$&$22.85\pm0.16$&$-2.02$&$ -0.20\pm0.28$&$-1.88$&$-0.05$\\
S04&$ -0.079$&$ -0.237$&$22.22\pm0.09$&$-2.65$&$ -0.86\pm0.25$&$-2.30$&$-0.49$\\
S05&$ -0.064$&$ -0.045$&$22.68\pm0.14$&$-2.19$&$\phantom{-}0.14\pm0.20$&$-2.06$&$\phantom{-}0.24$\\
S06&$ -0.061$&$  0.061$&$22.75\pm0.14$&$-2.12$&$ -0.50\pm0.30$&$-1.89$&$-0.25$\\
S07&$ -0.047$&$  0.016$&$21.67\pm0.06$&$-3.20$&$ -0.66\pm0.13$&$-2.92$&$-0.36$\\
S08&$ -0.014$&$ -0.158$&$23.06\pm0.19$&$-1.81$&$ -1.67\pm0.82$&$-1.16$&$-1.00$\\
S09&$ -0.005$&$  0.072$&$22.43\pm0.11$&$-2.44$&$ -1.29\pm0.40$&$-1.94$&$-0.77$\\
S10&$ -0.005$&$  0.128$&$23.27\pm0.23$&$-1.60$&$ -0.74\pm0.53$&$-1.29$&$-0.42$\\
S11&$  0.000$&$  0.000$&$21.05\pm0.03$&$-3.82$&$\phantom{-}0.27\pm0.04$&$-3.71$&$\phantom{-}0.36$\\
S12&$  0.005$&$ -0.092$&$22.66\pm0.13$&$-2.21$&$ -0.96\pm0.38$&$-1.82$&$-0.56$\\
S13&$  0.029$&$  0.235$&$22.93\pm0.17$&$-1.94$&$ -1.32\pm0.60$&$-1.42$&$-0.79$\\
S14&$  0.038$&$ -0.231$&$22.92\pm0.17$&$-1.95$&$ -0.66\pm0.39$&$-1.68$&$-0.36$\\
S15&$  0.052$&$  0.063$&$21.60\pm0.05$&$-3.27$&$ -0.64\pm0.12$&$-3.00$&$-0.35$\\
S16&$  0.103$&$  0.212$&$22.84\pm0.16$&$-2.03$&$ -1.01\pm0.46$&$-1.64$&$-0.59$\\
S17&$  0.114$&$  0.136$&$22.78\pm0.15$&$-2.09$&$ -0.58\pm0.33$&$-1.84$&$-0.31$\\
S18&$  0.120$&$  0.027$&$22.62\pm0.13$&$-2.25$&$ -0.27\pm0.24$&$-2.10$&$-0.09$\\
S19&$  0.169$&$  0.092$&$22.79\pm0.15$&$-2.08$&$ -0.93\pm0.42$&$-1.71$&$-0.54$\\
S20&$  0.177$&$ -0.005$&$22.24\pm0.09$&$-2.63$&$ -0.51\pm0.20$&$-2.40$&$-0.26$\\
\enddata
\label{tab:stars}
\tablecomments{The $m_{330}$ values are not corrected for galactic extinction,
while all values in the subsequent columns are corrected assuming 
$A_U=0.44$ mag, and $A_B=0.36$ mag.  Standard Johnson U and B values are
derived using the color transformations in \citet{acs}. The distance
to M31 is assumed to be 770 kpc \citep{dm31}. S07 and S11 are extended and
are probably blends of several stars.  S11 is hypothesized to be the location
of the M31 black hole.}
\end{deluxetable}

\clearpage

\end{document}